\documentclass[twocolumn,prb,showpacs]{revtex4}  
\def\be{\begin{equation}}
\def\ee{\end{equation}}
\def\beq{\begin{eqnarray}}
\def\eeq{\end{eqnarray}}

\usepackage{subfigure}
\usepackage{graphicx}
\usepackage{rotating}
\usepackage{amsmath}
\usepackage{amsfonts}
\usepackage{amssymb}
\usepackage{enumerate}
\usepackage{longtable}
\usepackage{dcolumn}
\usepackage{bm}
 
\begin{document}

\title{Mott transitions and Novel Orders in Multi-Orbital Models: The Relevance of Structural
``Double Exchange''}
 
\author{Mukul S. Laad$^1$, S. Koley$^2$ and A. Taraphder$^2$ }
 
\affiliation{$^{1}$ Institute of Mathermatical Sciences, Chennai 600013, India \\
$^{2}$Department of Physics and Centre for Theoretical studies,\\
Indian Institute of Technology, Kharagpur 721302 India} 
 
\date\today

\begin{abstract}
 
In real transition-metal oxides, the so-called GdFeO$_{3}$ octahedral
tilt is long known to be a relevant
control parameter influencing the range of orbital and magnetic ordered
states found across families of
cubic perovskite families. Their precise role in the interplay
between itinerance and localisation, long known
to underpin Mottness in $d$-band oxides, has however received much
less attention.  We analyse the relevance
of the GdFeO$_{3}$ tilt in detail in a representative setting of a
partially-filled $e_{g}$ orbital system. We
identify a new generalised principle of broad relevance, namely, that
this tilt acts like a structural ``double
exchange'' and acts contrary to the well-known Anderson-Hasegawa
double exchange.  As a function of this tilt,
therefore, a phase transition from a Mott-Hubbard insulator to an incoherent bad-metal occurs as an effective
band-width-controlled Mott transition.  We analyse the incoherent metal in detail by studying one- and two-particle
spectral responses and propose that this selective-metal is a novel orbital analogue of the FL$^{*}$ state with
fractionalised {\it orbitons}.  Finally, we apply these ideas to qualitatively discuss the effect of strain on thin
films of such $d$-band oxides on suitable substrates, and discuss the exciting possibility of engineering
novel ordered phases, such as unconventional circulating currents, nematics and superconductors, by suitable strain engineering in TMO thin films.

\end{abstract}

 
\maketitle

Mott transitions, induced either by bandwidth control or doping~\cite{imada} are by
now believed to underpin novel phenomena in transition metal oxides (TMO). Among many
others, occurence of spectacular phenomena such as unconventional high-T$_{c}$
superconductivity~\cite{pwa}, colossal magnetoresistance~\cite{dagotto} and
multiferroicity~\cite{sen}, along with huge changes in response to minute perturbations
open up attractive potential for designing new devices where these features can be
profitably exploited.  On the fundamental side, in multi-orbital systems, additional,
interesting physics related to {\it orbital} selective Mott transition(s) emerges
in the intermediate correlation regime~\cite{os}.  This generically leads to bad-metallic,
non-Landau Fermi liquid (nLFL) metallic behavior, topological restructuring of the
large Fermi surface (FS) of the LFL, and fractionalisation of elementary excitations~\cite{senthil},
as exemplified in the FL$^{*}$ idea.  Given the multi-band situation intrinsic to most
real oxides, these novel features may be more broadly found in TMOs than currently
known.  In particular, whether an analogous orbital-correlation driven orbital-FL$^{*}$
(i.e, OFL${*}$) state with fractionalised (charge neutral) fermionic orbitons can be
realised remains completely unexplored, despite proposals for orbitons as analogues of
magnons being extant in literature~\cite{jeroen}. 
 
In this context, recent proposals to use heterostructuring and orbital engineering in
concert with remarkable advances in synthesising artificial layered structures of
TMO have provided exciting impetus to attempts to design materials with exotic
properties not to be found in the bulk~\cite{takahashi}.  Various factors, e.g, quantum
confinement, interface orbital polarisation due to chemical confinement and epitaxial
constraint on crystal symmetry are believed to conspire in selecting specific electronic
ground states in superlattices of TMO. One expects such a wide range of ground states
to be selected by ``tuning'' the system, e.g, by external strain, which can be
applied by using appropriately terminated substrates, e.g, a 2.2 percent tensile
strain on LaNiO$_{3}$/LaAlO$_{3}$ (LNO/LAO) superlattices has been achieved~\cite{ref}
by using a TiO$_{2}$-terminated (001) SrTiO$_{3}$ substrate.   
 
  As regards orbital engineering, chemical confinement and epitaxial constraints will modify
the interfacial orbital character of electronic states in TMO films relative to their
bulk character. Tensile or compressive strain will also produce such changes~\cite{pd-atapl},
with the added attractive advantage of providing a knob to ``continuously'' tune the
orbital character. In TMO, this must occur as a result of changes in the GdFeO$_{3}$
tilt, known to be a ubiquitous feature. More precisely, in corner-sharing $TMO_{6}$
octahedra in TM oxides, off-equilibrium displacements of the lighter $O$ ions
induces coupled distortions into the problem.  This leads to anti-ferrodistortive
displacements, resulting from octahedral tilt, the so-called $GdFeO_{3}$ distortion.
In situations where the one-electron hopping between the TM ions is mediated by
the intervening $O$ ions, as in cuprates and manganites, this tilt directly affects
wavefunction overlap and degree of itinerance. Since the remarkable responses of
TMOs are a direct fall-out of the competition between itinerance and local interactions,
the resulting $GdFeO_{3}$ tilt angle, $\theta$, turns out to be a relevant ``control
parameter'', readily tunable by external pressure or uniaxial strain. In cubic $ABO_{3}$
perovskites, the $A$-atom size controls the magnitude of $\theta$, and is known from
many studies~\cite{tokura,goodenough} to be an important feature in evolution of
ground states and physical properties as a function of $A$. Clearly, this globally
important feature must also play an important role in evolution of ground states and
physical responses in TMO heterostructures. To the best of our knowledge, how this
specific aspect affects the competition between itinerance and Mott localisation in
TMO, along with the more fundamental theoretical issues mentioned above, remains to
be addressed systematically.
 
We start by searching for a basic theoretical systematization of the trends
found in the better studied cubic ABO$_{3}$ systems. Partly, difficulty in achieving
this stems from the paucity of well-controlled theoretical tools capable of describing
{\it all} effects (itinerance, local Hubbard interactions, Jahn-Teller (JT) distortions,
$GdFeO_{3}$ tilt) in {\it real} TMOs in a single picture. An aim of our study is to retain
essential aspects of the general problem in a simplified effective model, yet one which should be tractable
enough to give a broad applicability and predictive power.  
 
Here, we illustrate how such trends as found experimentally across a wide range of
TMO families are rationalized in such a ``simple'' effective model. Specifically, we consider
how major trends above vary with $\theta$. We show that a wide variety of such
trends in diverse TMO~\cite{palstra} can be understood directly in terms of a very
simple underlying theme, whose derivation and study of consequences is the main result
of this work:
 
{\it The GdFeO$_{3}$ tilt in TMOs acts as a ``structural double exchange'' in
the orbital sector.  It acts contrary to the well-known Anderson-Hasegawa
double exchange (DE), and the competition between these two double exchange processes
is at the heart of the unusually rich and complex outcomes observed in
experiment as structural features are appropriately tuned.}
 
Building upon these strengths, we apply this theme to study the Mott insulator-to-bad
metal transition in a single layer TMO (e.g, manganite) film described by a two-orbital
Hubbard model including ``realistic'' features specific to e$_{g}$-orbital systems.  Using an effective model
for the interface based on DMFT results, we also discuss how novel orders in TMO thin films on suitable substrates can arise via route(s) unanticipated in the case of bulk TMOs.  We
stress, however, that our model and results have a broad application to
TMOs in general, and our results admit an in-principle generalisation to t$_{2g}$-based TMOs as well.

\section {Model}
 
Specifically, motivated by intense interest in manganite and nickelate-based thin
films~\cite{stewart}, we consider a 2D, two-orbital ($e_{g}$ orbitals) system with
the $GdFeO_{3}$ octahedral distortion, with Hamiltonian,
 
\be
H=H_{band}+H_{int}+H_{JT}
\ee
 
\noindent where $H_{band}$ consists of the hopping part in the $e_{g}$ sector, plus the
$GdFeO_{3}$ distortion, and $H_{int}$ is the multi-orbital Coulomb part.
Explicitly, $H_{band}=-\sum_{<i,j>,a,b}t_{ij}^{ab}(a_{i\sigma}^{\dag}b_{j\sigma}+h.c)$
where the $t_{ij}^{ab}$ have very specific orbital dependent structures for the
two $e_{g}$ orbitals in $2D$ ($a,\,b$ stand for the two $e_g$ orbitals, e.g., $d_{x^2-y^2}$ 
and $d_{3z^2-r^2}$ for manganites). The $GdFeO_{3}$ distortion can be included in 
$t_{ij}^{ab}$ via an angle, $\theta$, describing the rotation of the $TMO_{6}$ octahedra.
We have $t^{aa}=(3t/4)cos^{3}\theta$, and $t^{bb}=(t/4)cos\theta$, while
$t^{ab}=t^{ba}=(\sqrt{3}t/4)cos^{2}\theta$ for hopping along $y$, while
$t^{ab}=t^{ba}=(-\sqrt{3}t/4)cos^{2}\theta$ for hopping along $x$.
Interestingly, the inter-orbital hopping has an explicit $d$-wave character
with a form-factor (cos$k_{x}$-cos$k_{y}$) in momentum space, a feature that
can facilitate unconventional orbital order, detailed later in this work.
 
The multi-orbital Coulomb interactions in the $d$-shell are described by
$H_{int}=U\sum_{i,\alpha=a,b}n_{i\alpha\uparrow}n_{i\alpha\downarrow} +
U'\sum_{i}n_{ia}n_{ib} - J_{H}\sum_{i}{\bf S}_{ia}.{\bf S}_{ib}$. Additionally, for
partially filled $e_{g}$ orbitals, the JT distortion lifts the $e_{g}$
orbital degeneracy at the outset, contributing the well-known term,
$H_{JT}=gQ_{2}\sum_{i\sigma}(a_{i\sigma}^{\dag}b_{i\sigma}+h.c) +
gQ_{3}\sum_{i\sigma}(n_{ia\sigma}-n_{ib\sigma})$, to the Hamiltonian. $Q_2$ and $Q_3$ are 
the relevant distortions of the lattice (see J. Kanamori~\cite{kanamori} for the standard  
notations and definitions of the Jahn-Teller distortion magnitudes $Q_{2},Q_{3}$). 
To simplify matters, we consider {\it quasi-static} JT distortions
to begin with: this seems to be quite adequate for the case of $D=2$ (bilayer) manganites, as
found very recently~\cite{boothroyd} but it is an approximation. In this limit, the JT term can be written in
a manifestly ``orbital double-exchange'' form~\cite{dagotto}: $H_{J-T}=
g\sum_{i}{\bf Q}_{i}.{\bf T}_{i}$ where ${\bf Q}=(Q_{2},Q_{3})$ and ${\bf T}=
(T^{x},T^{z})=((n_{a}-n_{b})/2, a^{\dag}b+b^{\dag}a)$.
 
The band Hamiltonian, $H_{band}$, is readily diagonalized: we introduce the combinations,
$c_{\sigma+}=\sum_{\alpha=e_{x},e_{y}}\frac{\sqrt{3}c_{\theta}^{3/2}.a_{\sigma}+c_{\theta}^{1/2}(-1)^{\alpha}b_{\sigma}}{\sqrt{c_{\theta}(1+3c_{\theta}^{2})}}$ and
$c_{\sigma-}=\sum_{\alpha=e_{x},e_{y}}\frac{(-1)^{\alpha}c_{\theta}^{1/2}.a_{\sigma}-\sqrt{3}c_{\theta}^{3/2}.b_{\sigma}}{\sqrt{c_{\theta}(1+3c_{\theta}^{2})}}$ 
with $c_{\theta}=$cos$\theta$. Here, $(-1)^{\alpha}=+1,-1$ for $\alpha=e_{x},e_{y}$. The $c_{\pm}$ now defines new, orthogonal combinations of $a,b$ in the $GdFeO_{3}$ structure. In $D=2$ manganites, with
$a=d_{x^{2}-y^{2}}, b=d_{3z^{2}-r^{2}}$, the $c_{\pm}$ transform 
like $d_{3x^{2}-r^{2}},d_{3y^{2}-r^{2}}$. Quite remarkably, geometric effects allow
only the $c_{\sigma+}$ to hop: the $c_{\sigma-}$ remains completely dispersionless
as long as no additional distortions are introduced: this feature was also apparently discovered earlier by Ferrari {\it et al.}~\cite{ferrari}.  In the limit $U\rightarrow\infty$,
(as in effective FM Kondo lattice models for manganites in the large-J$_{H}$ limit~\cite{furukawa})
the interaction terms can be easily re-expressed in terms of the $c_{\pm}$ as
$H_{int}=U'\sum_{i,\sigma,\sigma'}n_{i\sigma+}n_{i\sigma'-} -
J_{H}\sum_{i}{\bf S}_{i+}.{\bf S}_{i-}$ while the hopping part is simply
$H_{hop}=(t/4)\sum_{i,\alpha=x,y,\sigma}\gamma_{ij}(\theta_{ij})
(c_{i\sigma+}^{\dag}c_{i+\alpha,\sigma+}+h.c)$.
Results for the original two-band Hubbard model with $d$-wave hybridisation are readily
obtained by undoing the transformation from $a,b$ to $c_{+},c_{-}$ fermions.
 
Interestingly enough, this is just the $S=1/2$ Falicov-Kimball (FK) model, as long as
the JT term is neglected, and the hopping is uniform, i.e, when $\gamma_{ij}(\theta_{ij})
=C$, a constant. Otherwise, one has an FK model with bond-dependent
hopping, $t_{ij}=t\sqrt{c_{\theta_{i}}(1+3c_{\theta_{i}}^{2})}\sqrt{c_{\theta_{j}}
(1+3c_{\theta_{j}}^{2})}$. Following this, incorporating a Holstein-like electron-lattice
$(e-l)$ interaction in our case will obviously further enhance the
FK-like nature of our model: the dispersionless $c_{\sigma,-}$ states above will now
favor polaron formation, even for much more modest e-l coupling. The dispersive band
($c_{\sigma,+}$) will be moderately narrowed by electron-lattice interaction, and this
effect can be subsumed into a renormalised hopping strength in $H$ above. In other words,
we do {\it not} need strong electron-lattice coupling, in accord with experimental
indications for cubic manganites~\cite{mathur}. Geometrical effects leading
to drastic reduction in band-width for a subset of orbital states will allow polaron
formation even for modsest $e-l$ coupling.
 
The new element in our work is that the one-electron hopping explicitly shows up
in the dependence on the GdFeO$_{3}$ tilt ($\theta$) in a simple way. Variation
of $\theta$ will thus lead to a (continuous) variation of $U/t(\theta),U'/t(\theta)$ in
our FKM, whence the exciting new possibility of a Mott transition driven by octahedral
tilt reveals itself in the form of a bandwidth-controlled Mott transition. In the rest
of this work, we will use multi-orbital DMFT to solve $H$ and discuss the interesting
manifestations of this effect in the cases outlined in the introduction.
 
\section {$GdFeO_{3}$-Tilt-induced Mott transition}
 
Many interesting trends in ground states and physical responses are now readily manifest.
To this end, it is helpful to look at a generic situation, which we do by replacing the
actual tight-binding DOS for the $e_{g}$ orbitals above by a simple Bethe lattice DOS.
It is known~\cite{voll} that the FKM shows a continuous Mott (metal-insulator) transition as $U/zt(\theta)$
is varied through a critical value. For the spinless FKM (corresponding to the saturated
half-metal with full ferromagnetic spin polarization due to an effective
$J_{H}=\infty$~\cite{furukawa}) on the Bethe lattice, the critical $U'$ is $U'=
2\sqrt{3}zt(\theta)$ with $t=t(\theta)=t[c_{\theta}(1+3c_{\theta}^{2})]$, for $\theta_{i}=-\theta_{j}$, corresponding to antiferrodistortive octahedral tilt commonly expected in many TMOs.
 
Thus, for fixed $U'$, increasing the $GdFeO_{3}$ tilt {\it reduces}
$W(\theta)$, driving a Mott metal-insulator transition as a function of the
{\it tilt angle}, $\theta$!  This can readily be induced by applying pressure, or
from coupling to substrate-induced {\it strain} (see below).
Increasing $\theta$ also reduces the AF transition temperature: remarkably, at strong
coupling, $k_{B}T_{N}^{af}\simeq (4t^{2}/U)$ decreases in tandem with
{\it stabilization} of the Mott insulator. At weak-coupling, $k_{B}T_{N}\simeq
t(\theta)$exp$[-2\pi(\theta)/U)^{1/D}]$ within HF-RPA, giving a qualitatively different
scaling of $T_{N}(\theta)$. The orbital ordering scales also follow similar laws, 
$k_{B}T_{oo}=(4t^{2}/U')$ for large $U'$ and $k_{B}T_{oo}\simeq
t(\theta)$exp$[-2\pi(\theta)/U')^{1/D}]$ for small $U'$. We thus suggest that studying
the pressure or strain dependence of $T_{N}$ and $T_{oo}$ (by varying the $A$ cation in
ABO$_{3}$ perovskites) can yield direct information concerning the degree of electronic
correlations (large or small $U$) and the itinerant vis-a-vis Mottness nature of magnetism
in $d$-band magnetic oxides.
 
The above is very simply understood by noticing that the GdFeO$_{3}$ tilt acts like
a ``reverse double exchange'' - increasing $\theta$ reduces $t_{ij}(\theta)$, disfavoring
the metallic state and ferromagnetism, but favoring AF-OO Mott insulators. 
Also, in half-metals, it now turns out that incorporating this {\it structural} DE into the
conventional Anderson-Hasegawa~\cite{anderson} argument gives
$t_{ij}=t(\theta_{ij})\sqrt{1+(<{\bf S}_{i}.{\bf S}_{j}>/2S^{2})}$, revealing the interplay
between the structural (reverse, $t(\theta_{ij})$) and conventional DE ($\sqrt{1+(<{\bf S}_{i}.{\bf S}_{j}>/2S^{2})}$) in a very transparent way.  Incorporation
of the all-pervasive structural (GdFeO$_{3}$) tilt as a generalized principle, i.e, a ``reverse''
DE has not been previously realised, to our best knowledge.  The major technincal advantage
of our analysis is thus that such effects can now be easily incorporated into modern
correlated electronic structure (LDA+DMFT) approaches as multi-orbital Hubbard models with
bond-dependent hoppings.  
 
Clearly, {\it decreasing} tilt now favors the FM-metal, self-consistently {\it reducing} the
JT distortion via enhanced itinerance.  Increasing $\theta$ counteracts this
tendency, reducing the hopping, favoring OO and JT distorted
across the $AMO_{3}$ with $M=Ti,Ni$, one should find AF Mott insulators (large $\theta$)
giving way to correlated metals (smaller $\theta$) as $A$ is varied: indeed,  this fully
accords with data, where precisely such a correlation between the magnitude of the
GdFeO$_{3}$ tilt and evolution of electrical properties (metal or insulator)
is known~\cite{goodenough,tokura}. Further, the ``puzzling'' weakening of AF order
along with stabilisation of the Mott insulator in $RNiO_{3}$ also falls out rather
naturally as discussed above. We emphasise that, in the above, the effective parameters,
$t(\theta), U'$ in our FKM include various structural (one-electron bands), chemical (in
the effective value of $t$, reflecting strong covalent $p-d$ overlap in late
TMO via $t\simeq t_{pd}^{2}/\Delta$ where $t_{pd}$ is the one-electron hybridization
and $\Delta=(\epsilon_{p}-\epsilon_{d})$ is the charge transfer energy) as well as
interaction ($U,U',J_{H}$ and JT couplings) in a simple effective model. This enabled
us to understand global trends in evolution of ordered ground states and electrical
properties without explicitly referring to material-dependent microscopic
band structural details of specific TMO with partially-filled $e_{g}$ orbitals.
Thus, our derivation of the structural DE and conclusions following from its interplay
with conventional DE have a more general validity, as exemplified by good qualitative
agreement discussed above.
 
Thus, we identify the competition between the structural ``reverse'' DE and conventional DE
as the central factor influencing the range of tunable ground states in cubic TMO.
However, much work remains to be carried out when one inquires about the role of the
above competition in strained thin films (or devices made from sandwiches of such films)
involving $ABO_{3}$ perovskites, e.g LSMO/STO~\cite{hrk} or LNO/LAO
interfaces~\cite{chakhalian}.
 
\section {Application to strain-induced Mott transition in TMO thin films}
 
We now apply the ideas discussed in the introduction and developed in the preceding
sections to a very contemporary problem in TMO thin films. Namely, we will consider
the following issue, albeit with a simplified model band structure:
 
Imagine a situation where an atomically thick layer of a TMO, e.g, a manganite or a
nickelate, deposited on an appropriately terminated substrate, e.g, (001) face
of SrTiO$_{3}$ or LaAlO$_{3}$ (present advances in technology make this exciting situation
possible, and, moreover, the thickness can be varied between one and several TMO layers).
In this situation, several new features, not important for bulk TMO, become relevant. The
excitement in this field stems from the belief that these allow for additional ``knobs''
to tune the interplay between various microscopic physical processes, and open the door
to engineering desired ground states, which can be further manipulated by small changes in
external parameters, i.e, {\it by slowly turning the knob}. We list these features:
 
(i) Depending upon the substrate, interfacial strain can be manipulated to be either
tensile or compressive. In an $e_{g}$-orbital based system, this modifies the multi-orbital
one-electron hoppings, as well as the Jahn-Teller splittings. Assuming a (001)
growth direction and defining the inlayer strain by $e_{xy}=(a_{s}-a)/a_{s}$ with $a,
\, a_{s}$ the lattice constants of the manganite film and the substrate, $e_{xy}>0\,(<0)$
corresponds to tensile (compressive) strain. The strain along $c$-axis, $e_{z}=-
4\nu e_{xy}$ with $\nu$ being the Poisson ratio. For manganites, we choose $e_{z}=-
3e_{xy}/2$ and $-0.02\leq e_{xy}\leq 0.02$.  This is the same strategy as that employed recently
by Baena {\it et al.}~\cite{baena}, and has a number of interesting consequences relevant to
our study: 
 
\noindent (a) strain modifies the nearest neighbor hopping integral as
$t=t(1-2e_{xy})$, leading
to reduction (enhancement) of itinerance for tensile (compressive) strain.
 
\noindent (b) Tensile strain lowers the on-site energy of the $x^{2}-y^{2}$ orbital
($\epsilon_{1}$) relative to that of $3z^{2}-r^{2}$, denoted $\epsilon_{2}$, while
compressive strain does the opposite, i.e, it effectively modifies the magnitude
{\it and} sign of the JT distortion. This effect is modelled by a term,
$\sum_{i,\sigma}(\epsilon_{1}n_{i,a,\sigma}-\epsilon_{2}n_{i,b,\sigma})$ in our
original Hamiltonian.  Written in terms of the $c_{\sigma,\pm}$, it is exactly of the 
form of $H_{JT}$. The important difference is that the {\it sign} of this {\it strain
induced} ``JT''-like distortion can be opposite to that of the usual JT coupling. It is
also known that the magnitude of the GdFeO$_{3}$ tilt controls (for example, by choosing
the $A$
atom in AMnO$_{3}$) the magnitude of the usual JT interaction, even for bulk $D=3$
manganites: the same must then hold for manganite films as well. 
Also, epitaxial constraints at the interface must lead to octahedral tilts ($\theta$)
notably different from their values in the bulk. As we have seen above, this acts like
a ``reverse'' DE, and, by itself, affects the electronic bandwidths and Jahn-Teller
splittings, as discussed in detail for the case of bulk AMnO$_{3}$.
The points (a) and (b) above clearly illustrate how both, the one-electron hoppings
{\it and} the {\it net} value of the JT coupling, are rather sensitive to the combined
effects of the GdFeO$_{3}$ tilt and strain.   
 
\noindent (ii) Strong local multi-orbital Coulomb interactions are known to be ubiquitous in
TMO, and correlation effects in thin layers are known to be generically enhanced compared to
the bulk, due to reduction of screening.
 
It is very natural to expect that all these features will play an important role
when one studies the ground state(s) and associated phase transition(s) between
them as tuning parameters like strain or composition are varied. We will consider
TMO thin films (e.g, manganite layer on STO substrate) in some detail in this context in what
follows. We consider the more realistic {\it spinful} two-orbital Hubbard model defined in
the previous section. Our main focus will be on how the electronic properties of
the layer can be sensitively switched by strain engineering,
and, in general, by any perturbation which affects the interplay between structural
(GdFeO$_{3}$ tilt, JT coupling), itinerant (one-electron hoppings via GdFeO$_{3}$ tilt)
and strong multi-orbital Hubbard interactions. We will also qualitatively discuss their
effects on the relative stability of competing magnetic and orbital ordered states, and
the broader relevance of our analysis here to other interesting issues
that have been the recent focus of attention in TMO heterostructures and thin films.
 
The two-band Hubbard model for a single-layer, written in terms of
the $c_{\sigma,\pm}$ combinations, becomes the spin-$1/2$ FKM with an additional
JT term
 
\begin{equation*}
\begin{split}
H_{2D}=\sum_{<i,j>,\sigma}t\gamma_{ij}({\bf S})\gamma(\theta_{ij})
(c_{i,\sigma,+}^{\dag}c_{j,\sigma,+}+h.c)+U\\
\sum_{i,a=\pm}n_{i,a,\uparrow}n_{i,a,\downarrow}+
U'\sum_{i,\sigma,\sigma'}n_{i,\sigma,+}n_{i,\sigma',-}
+ \lambda\sum_{i}{\bf Q}_{i}.{\bf T}_{i},
\end{split}
\end{equation*}
as described earlier. The JT coupling is taken to be $\lambda$ and not $g$, as before,
to signify that this quantity will be renormalised by strain, as discussed above. 
We re-emphasise that, in our effective $S=1/2$ FKM, even a modest electron-lattice
interaction now drives the system to the anti-adiabatic regime ($\Omega_{ph}/t_{-} \gg 1$)
in the $c_{-}$-sector, while, due to a wide band width ($O(2.0)$~eV) of the $c_{+}$,
the adiabatic limit is applicable in the latter case. This is intrinsically an orbital-selective
electron-lattice coupling, a new feature
rigorously demonstrated in our case at the level of a tight-binding band structure
(the fact that a very similar tight-binding fit, though with different features, works
very well for cubic manganites, at least in the FM metallic phase, was shown earlier
and utilised in a DMFT calculation~\cite{laadman}). Our new effective picture is thus of
immobile small polarons in $c_{\sigma,-}$ states co-existing with comparatively wide-band dispersive
$c_{\sigma,+}$ electronic states. For the wide ($c_{+}$)-sector, the adiabatic limit
implies that we may consider the effect of the $\lambda\sum_{i}{\bf Q}_{i}.{\bf T}_{i}$
term within the quasi-static approximation (neglecting $\Omega_{ph}$, since, in this limit,
the Holstein problem reduces to a separate Falicov-Kimball model~\cite{freericks}), and this
additional FK interaction, $H_{FK}'=u'\sum_{i,\pm}\Delta_{i}n_{i,c,\pm}$ can be lumped together
with the $U'\sum_{i,\sigma,\sigma'}n_{i,\sigma,+}n_{i,\sigma',-}$ in $H_{2D}$ above. 
 
Thus, we are now left with having to analyse the Hamiltonian $H_{2D}$.  Due to the fact
that the term $\lambda\sum_{i}{\bf Q}_{i}.{\bf T}_{i}$ mixes the $c_{+},c_{-}$ states
at one-electron level, we have to solve a two-orbital extended Anderson lattice model.
To study the questions posed in this work, we solve $H_{2D}$ within multi-orbital
dynamical mean field theory (MO-DMFT), using the MO-iterated perturbation theory (MO-IPT)
as the impurity solver in DMFT.  A wide range of impurity solvers can be used in this
context: these range from diagrammatic ones like IPT~\cite{laad}, non-crossing
approximation (NCA)~\cite{kot1}, local moment approximation (LMA~\cite{logan}, to
numerical exact diagonalisation (ED)~\cite{liebsch}, continuous-time quantum Monte Carlo
(CT-QMC)~\cite{millis} and numerical renormalization group (NRG)~\cite{krish}. Both
approaches yield the correct correlated Landau Fermi liquid (LFL) correlated metallic state
and first-order Mott transition in the one-band Hubbard model as well as the selective-Mott
transitions accompanied by bad-metallic incoherent non-LFL metallic states in correlated
multi-band systems~\cite{kot-rev,held}, though, to our klnowledge, only NRG achieves the
necessary accuracy required to extract the (almost) exact Kondo scale. Diagrammatic
approaches have the advantage of being relatively easy to implement, and work at all
temperatures: these advantages are not shared by the more exact purely numerical
approaches. In view of the ability of MO-IPT to access {\it both}, the selective-Mott
incoherent metal and the Mott transition, we use it in this work. The technical details
have already been worked out in detail in previous work~\cite{la-f}, and we refer
those interested in the details to these papers.
 
\section {Results}
 
We now describe our results. We choose $t$ to have a non-interacting bandwidth $W=3.0$~eV
for the $c_{+}$ fermions.  As mentioned, we denote the {\it net} JT coupling strength
by $\lambda$.  Assuming quasi-static JT distortions, the JT-term in $H_{2D}$ is quadratic
in the $c_{\pm}$, and is diagonalised along with the free part of the band structure to
yield two bands   
 
\be
E_{+}({\bf k})=\epsilon_{\bf k}+\sqrt{(\epsilon_{\bf k}+2q_{1})^{2}+4q_{2}^{2}}
\ee
 
\noindent and
 
\be
E_{-}({\bf k})=\epsilon_{\bf k}-\sqrt{(\epsilon_{\bf k}+2q_{1})^{2}+4q_{2}^{2}}
\ee
 
\noindent  where $q_{1},\, q_{2}$ are parameters chosen to be $0.05,0.05$~eV
as representative values, with the understanding that their {\it renormalised}
(by multi-orbital correlations and GdFeO$_{3}$ tilt) values will be very different,
as we will show by explicit computation below. The local Hubbard $U$ is chosen to be
$3.0$~eV, so that, with a $J_{H}=0.7$~eV, we have $U'\simeq (U-2J_{H})=1.7$~eV.
The hopping strength is varied either by varying the GdFeO$_{3}$ tilt angle ($\theta$)
or strain.  We keep in mind that, for the hopping between planar $x^{2}-y^{2}$
orbital states, tensile strain is ``equivalent'' to enhanced octahedral tilt
and reduced itinerance, while compressive strain corresponds to reduced tilt and
enhancement of itinerance. Thus, as predicted in earlier sections, a strain-induced
insulator-metal transition can be driven as a bandwidth-controlled Mott transition.
Investigation into details of this Mott transition, however, throw up surprises
we describe below in detail.
 
\begin{figure}[h]
\centering
\subfigure
{\includegraphics[angle=270,width=\columnwidth]{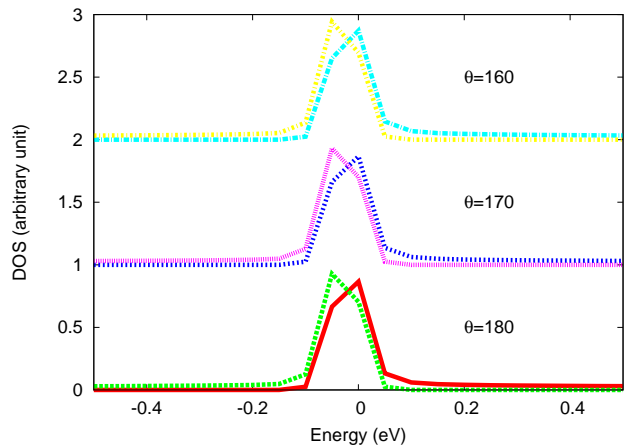}}
\caption{(Color Online) Non-interacting DOS of the two bands (see text) for three different tilt angles $\theta$ mentioned in the figure. Red, blue and cyan represent the DOS of the $E_{+}({\bf k})$ band at the three tilt angles shown and green, violet and yellow stand for the DOS of the $E_{-}({\bf k})$ band at the same tilt angles respectively.}
\label{fig1}
\end{figure}
\begin{figure}[h]
\centering
\subfigure
{\includegraphics[angle=270,width=\columnwidth]{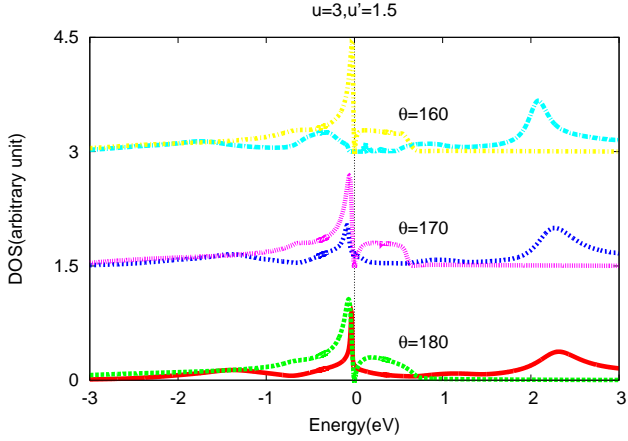}}
\caption{(Color Online) DMFT DOS of the two bands for three different tilt angles $\theta$, with intra-orbital Coulomb interaction $U=3$, and inter-orbital Coulomb interaction U$_{ab}$ =1.5. The color codes of the DOS are the same as in Fig.~\ref{fig1}.}
\label{fig2}
\end{figure}
 
\begin{figure}[h]
\centering
\subfigure
{\includegraphics[angle=270,width=\columnwidth]{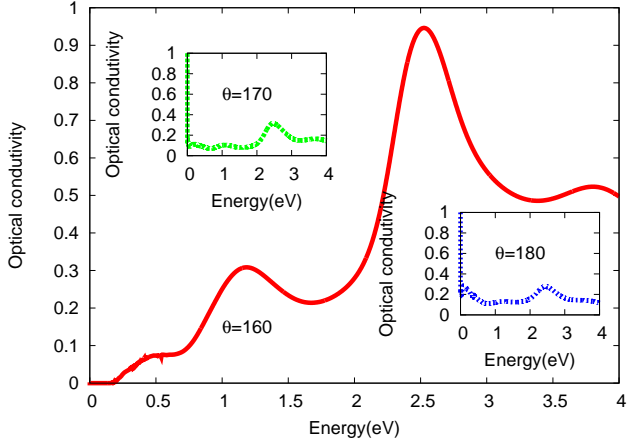}}
\caption{(Color Online) Optical conductivity from DMFT calculations for different tilt angles $\theta$ (main panel $\theta$=160$^{\circ}$, right inset $\theta$=180$^{\circ}$, and left inset $\theta$=170$^{\circ}$).}
\label{fig3}
\end{figure}
\begin{figure}[h]
\centering
\includegraphics[angle=270,width=\columnwidth]{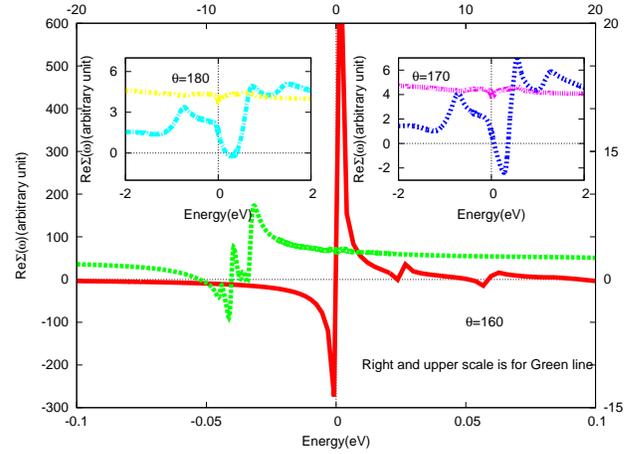}
\caption{(Color Online) (i) Real part of the DMFT self-energy at three different tilt angles (main panel $\theta$=160$^{\circ}$, right inset $\theta$=170$^{\circ}$, and left inset $\theta$=180$^{\circ}$). The different colors represent different bands as in Fig.~\ref{fig1}.}
\label{fig4}
\end{figure}

 \begin{figure}[h]
\centering
\subfigure
{\includegraphics[angle=270,width=0.8\columnwidth]{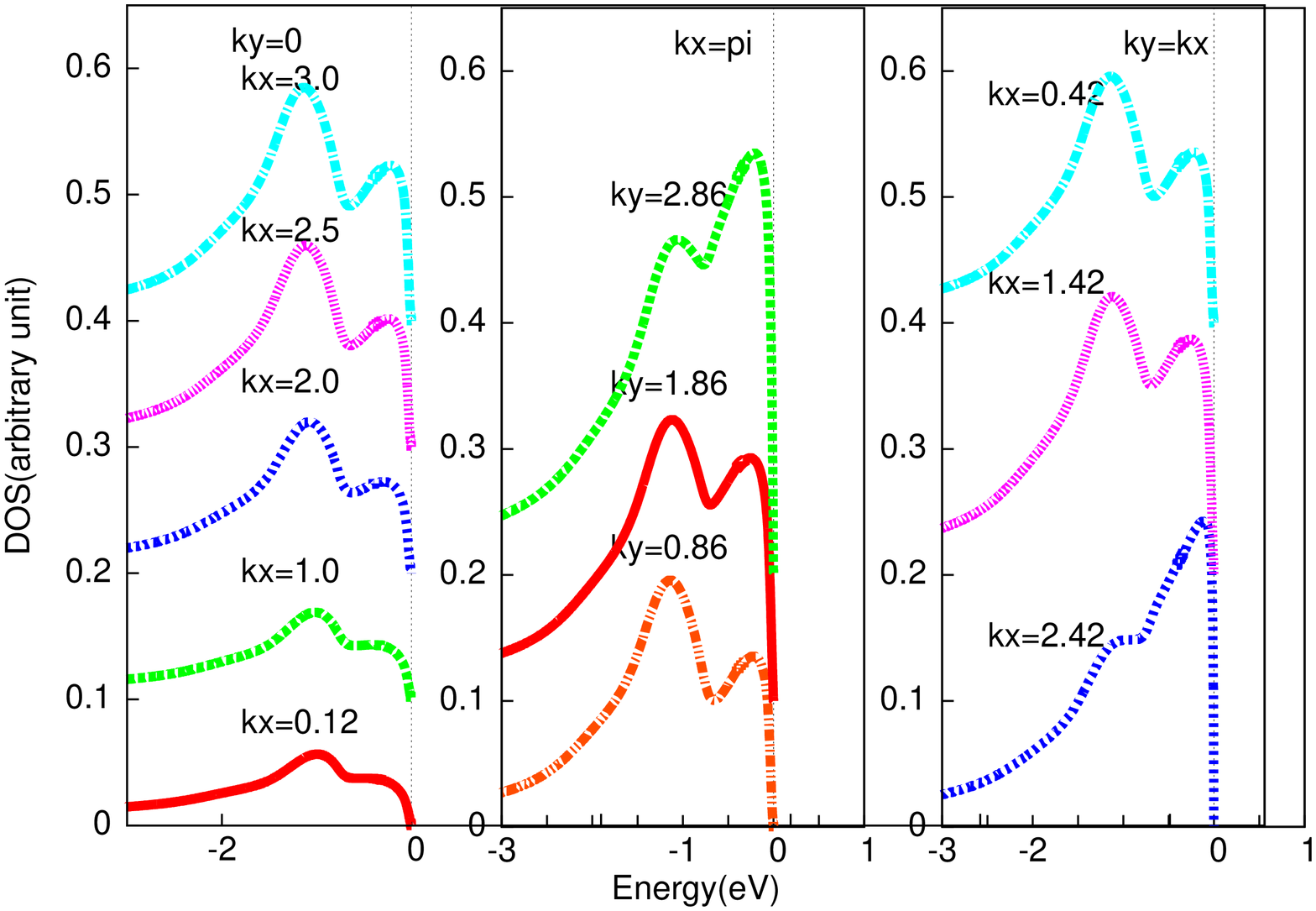}
}
{\includegraphics[angle=270,width=0.8\columnwidth]{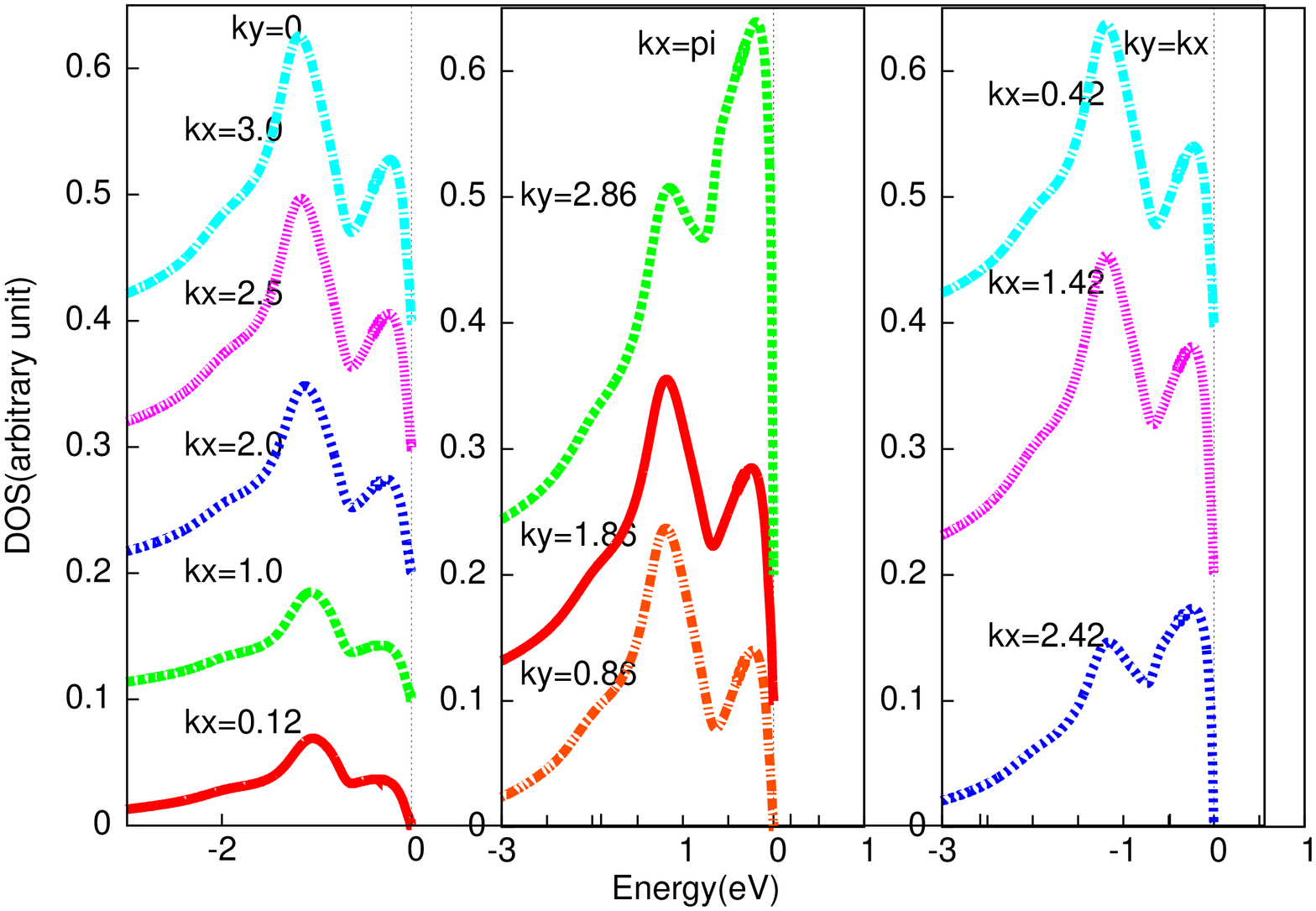}
}
\caption{(Color Online) The ARPES intensity at two different tilt angles (i) 180$^{\circ}$ and (ii) 170$^{\circ}$, calculated along different directions of k$_x$ and k$_y$ in the Brillouin zone as shown.}
\label{fig5}
\end{figure}

In Fig.~\ref{fig1}, we show the tight-binding band (TBB) DOS with two bands
$E_{+}(k,\theta),\, E_{-}(k,\theta)$, and the corresponding one-electron DOS in
the $c_{\pm}$ basis as a function of $\theta$, wherein the direct link between
itinerance and octahedral tilt is manifest as a reduction of the TBB width with
increasing tilt. The Fermi energy ($E_{F}$) is chosen to be at $\omega=0$, defining a
total band-filling, $n_{t}=n_{c,+}+n_{c,-}$.  It is clear that the van-Hove
singular (vHs) feature of the $c_{+}$-fermion band is pinned to $E_{F}$, hinting
at an intrinsic tendency toward instability to orbital density-wave (ODW) order,
which could then {\it co-exist} with ferromagnetism. This could be reasoned within
a weak-coupling, itinerant limit (small $U,\, U'$), as well as in the strong
coupling regime. However, observation of half-metallic behavior, as in the CMR
manganites, constrains one to use the latter route, as the strong $J_{H}(>W)$,
implies evel larger $U'U'$ from $d$-shell quantum chemistry. Generic observation of
insulating or bad-metallic states in such systems is
additional evidence for strong coupling physics. In this latter limit, one expects
increasing importance of coupled orbital-spin superexchange processes, rather
than the vHs, for instabilities toward ODW states. Finally, it is important to
notice that the $d$-wave inter-band hybridization (in the original ($a,b$) fermions)
is also reduced with increasing GdFeO$_{3}$ tilt: this will turn out to be useful later.
 
We now describe the effects of sizable local electronic correlations in detail.
In Fig.~\ref{fig2}, we show the orbital-resolved (DMFT) many-body density-of-states (DOS)
as a function of $\theta$. Several interesting features stand out clearly:
 
\noindent (i) for maximal $\theta=180$, we find metallic behavior, though without
LFL quasiparticles.
Increasing GdFeO$_3$ tilt (reducing $\theta$) reduces $t$, favoring increased (high-energy)
incoherence, and the bad-metal undergoes a {\it continuous} Mott transition to
a Mott-Hubbard insulator for $\theta=160$. As surmised above, this is a band-width
controlled Mott transition. Starting from the Mott insulator, increasing pressure now
generically reduces $\theta$, driving a continuous Mott transition to a bad-metal
accompanied by reduction of the {\it renormalised} JT-distortion. Quite interestingly,
precisely this is known to occur in LaMnO$_3$ under high pressure, and has
been investigated by Held {\it et al.}~\cite{held} within LDA+DMFT. We predict that
a simlar transition should obtain in bilayer manganites (or thin films) under
suitable ``pressure''.
 
\noindent (ii) In sharp contrast with the cubic manganites, however, the low-$T$
(FM in the DE case with large $J_{H}$ for mono- or bilayer manganites~\cite{tmatepl}) metallic state we find has
no stable LFL quasiparticles. Rather, our DMFT spectra
clearly show a low-energy continuum extending up to high energies $O(U)$.
This means that the Green functions show branch-cut behavior, rather than a
renormalised pole structure, in the complex energy plane. Remarkably, this non-LFL behavior
(of the $c_{+}$-band DOS) goes in tandem with a clear and strong tendency to (selective)
Mott localisation of the $c_{-}$ band (this is most clearly seen for the $\theta=170$
result in Fig.~\ref{fig2}). Examination of the orbital-resolved self-energies sheds more
light on this phenomenon, as shown in Fig.~\ref{fig3}: for the metallic cases ($\theta=170,
\, 180$), Re$\Sigma_{+}(\omega)$ shows sizably enhanced values near $E_{F}(=0)$, implying
heavily renormalised LFL quasiparticles, while Re$\Sigma_{-}(\omega)$ shows a clear
kink at $E_{F}$, implying divergent effective mass. Correspondingly, Im$\Sigma_{-}(\omega)
\simeq \omega^{(1-\eta)}$ with $\eta > 0$ at low energy, invalidating the LFL description
of the metal. In the Mott insulator, Im$\Sigma_{-}(\omega)$ clearly shows the (expected)
pole at $\omega=0(=E_{F})$.
Thus, our DMFT results describe an effective ``two-fluid'' situation, wherein Mott
-localized $c_{-}$ orbital states co-exist with (incoherent) ``metallic'' $c_{+}$
orbital states in the bad metal. This implies that though ODW-like states can be found
in weak-coupling descriptions, the selective-Mott metal lies outside their scope, and
only exists in the sizable correlation regime. Finally, the anomalous behavior of the
propagators and self-energies can only arise as consequences of strong {\it dynamical}
correlations, and are out of scope of LDA+U-type approximations.
 
\noindent (iii) In Fig.~\ref{fig5}, we exhibit the DMFT one-particle spectral function, $A({\bf k},\omega)=(-1/\pi)$Im$G({|bf k},\omega)$ for two values of the GdFeO$_{3}$ tilt angle.  The extinction of the quasi-coherent
LFL pole in the one-particle Green function directly translates into $(i)$ a power-law suppression in ARPES intensity as $\omega\rightarrow 0$.  Due to the $d$-wave form factor of the inter-orbital hybridization (which enters the unperturbed band structure of the two-band model), this suppression is also ${\bf k}$-dependent: ARPES lineshapes show maximal suppression for ${\bf k}$ along $(0,0)-(0,\pi)$ and clear $k$-dependent anisotropy.  This is a direct consequence of the complete suppression of the quasiparticle weight ($z_{FL}=0$),
as is clear from appearance of non-analyticities in Re$\Sigma(\omega=0)$ (see Fig.~\ref{fig4}).  The fact that
$G_{-}(k,\omega)$ now has {\it zeros} instead of poles at $\omega=0$ also implies that the unperturbed Fermi surface undergoes a non-perturbative reconstruction: only the $c_{+}$-fermions now contribute to the Luttinger
count.  $(ii)$ As the GdFeO$_{3}$ tilt angle is reduced, a clear transfer of spectral weight to high energies shows up as appearance of a marked shoulder around $\Omega\simeq -2.0$~eV for $\theta=170$ but not for $\theta=180$.  Its lack of momentum dependence, in strong contrast to that of the low-energy part, is instructive: reducing $\theta$ increases $U/t(\theta)$, transferring low-energy spectral weight to higher
energy which, clearly, must be associated with dispersionless Hubbard band-like states.  These observations are interesting in the context of studying latent electronic instabilities of TMO thin films: even in cases where the corresponding bulk systems show ``uninteresting'' paramagnetic metallic behavior~\cite{rotenberg}, it may be possible to generate even analogues of high-T$_{c}$ cuprates for thin films by appropriate strain tuning.  Explicitly, metallic LaNiO$_{3}$ thin films with appropriate misfit strain have been theoretically predicted to lie close to Fermi surface instabilities.  It is very conceivable that increase in effective $U/t$ via manipulation of the tilt angle due to epitaxial constraints can produce an OSMT.  If the reconstructed Fermi surface shows features propitious for electronic instabilities, a range of unconventional order(s) could result.
Such a ``strange'' metallic phase as we find here, with associated topological Fermi surface reconstruction(s), would be quite interesting, since it would open up avenues for observation of exotic concepts like fractionalized Fermi liquid phases in TMO thin film and superlattices context.

\noindent (iv) The microscopic origin of the bad metallic behavior is revealed upon
closer analysis: once the $c_{-}$-fermions are Mott-localized in DMFT, the effective
problem is that of the $c_{+}$-fermions scattering off Mott-localized $c_{-}$-fermion states.
In the local ``impurity'' problem in DMFT, this maps {\it exactly} to the inverse of the seminal
Anderson-Nozieres-de Dominicis x-ray-edge problem (this has been first recognized by Anderson in the context of the strange-metal phase in cuprates), wherein the infra-red
power-law singular feature arises due to the ``inverse'' orthogonality catastrophe (OC). 
Now, Im$G_{-}(\omega)\simeq \theta(-\omega)|\omega|^{-(1-\eta)}$, in full accord
with DMFT results. Identification with the X-ray edge problem has further interesting
consequences. It implies that the local inter-orbital ``excitonic'' susceptibility
also has infra-red power-law continuum behavior:
 
\begin{equation*}
\begin{split}
Im \chi_{+,-}(\omega)=\int d\tau e^{i\omega\tau} \langle c_{\sigma,+}^{\dag}c_{\sigma,-}(\tau);c_{\sigma,-}^{\dag}c_{\sigma,+}(0)\rangle \\
\simeq \theta(-\omega)|\omega|^{-(2\eta-\eta^{2})}
\end{split}
\end{equation*}
 
\noindent Hence, in the selective metal, orbital-excitons are not well-defined elementary
excitations (as they would be if the metal were a LFL). Instead of the conventional
long-lived particle-hole modes (i.e, a low energy coherent plasmon plus shake-up features at higher
energy in the LFL), the low-energy spectrum is characteristic of a critical orbital
{\it liquid}, in analogy with characterisation of spin-liquids: the particle-hole
spectrum is a critical (power-law) continuum of ``exciton-like'' origin extending up
to high energy. Thus, the (bosonic at long length scales) inter-orbital collective
plasmon is infra-red unstable in the non-LFL metal, and the continuum feature in
$\chi_{+,-}(\omega)$ strongly suggests charge-neutral {\it fermion-like} orbiton
excitations instead: this is the closest we can get to an orbital liquid with fermion-like
elementary excitations within DMFT. In fact, our finding
of a selective-Mott metal is very similar to that of the FL$^{*}$ fractionalised liquid
found in the context of $f$-electron metals by Senthil {\it et al.}~\cite{senthil}. In
our case, the renormalised Fermi surface (FS) of the metallic orbital liquid now has
a sharply defined sheet corresponding to the $c_{+}$ fermion band, while the $c_{-}$ FS
is ``critical''. We thus dub our metal as an orbital FL$^{*}$ (OFL$^{*}$) state, with
asymptotically decoupled charge and orbiton excitations at low energy. This is very
different from works where an orbital liquid state was proposed earlier on basis
of slave-boson Hartree-Fock theories~\cite{giniyat}, where computing higher-order
effects (beyond HF) of strongly coupled fermion-boson fluctuations present a formidable
problem (in the absence of these, difficulties similar to those found in slave-particle approaches to the $t-J$ model are expected to arise). As in the FL$^{*}$ case, we expect that the OFL$^{*}$ state is stable at low but intermediate energy scales in $D=2$, and instabilities to orbital-cum-magnetic ordered or LFL states will intervene at lower $T$ to relieve its extensive degeneracy (finite entropy per site). Thus, ordered
state(s) will arise directly from the non-LFL (``local'' critical) metal, rather than
from a band FS nesting-induced transition of a LFL, as would happen in the weak-coupling
limit.

   Onset of the OSMT also drastically transforms the bare band structure.  In the correlated LFL (non-OSMT) phase of the two-band model, adiabatic continuity with the free Fermi gas implies that the traditional perturbative version of Luttinger's theorem must hold.  The renormalized Fermi surfaces are then identical
(in shape and size) to that of the unperturbed two-band model, since the self-energies have no momentum dependence in DMFT.  Once the OSMT occurs, however, appearance of the $\omega=0$ pole in Im$\Sigma_{-}(\omega)$
implies that Im$G_{--}(k,\omega)=0$ at $k=k_{F},\omega=0$.  This causes a non-perturbative breakdown of
adiabatic continuity, leading to wipe-out of LFL quasiparticles in the OFL$^{*}$ phase~\cite{senthil}.  The
resulting ``Fermi surface'' is now a critical one, corresponding to a modified Luttinger count of $c_{+}$-fermions.  An important consequence of this non-perturbative (topological, driven by wipe-out of the $c_{-}$ Fermi surface by penetration of zeros of $G_{--}(k,\omega)$ in the OSMT phase to $\omega=0$) Fermi surface reconstruction is that instabilities to orbital- and/or magnetically ordered phases can no longer be rationalized by appealing to possible nesting features of the unperturbed Fermi surface(s), since the one-to-one
correspondence between the unperturbed and renormalized Fermi surfaces has been wiped out by onset of the OSMT.
As increasingly appreciated~\cite{phillips}, it is thus impossible to write down a traditional Ginzburg-Landau
field theory by integrating out the {\it unperturbed} fermions, since these do not exist as stable elementary excitations any more.  In fact, it is even impossible to write down a Luttinger-Ward functional when Im$\Sigma_{-}(\omega=0)$ diverges, emphasizing that any instability to ordered states from such an anomalous metal must be truly unconventional.
 
The above Mott transition and OFL$^{*}$ state leave their characteristic fingerprint on
one- and two-particle physical responses. In addition to canomalous continuum features in angle-resolved
photoemission (ARPES), optical conductivity studies
constitute a well-known spectroscopic diagnostic of two-particle response.
In Fig.~\ref{fig4}, we show the optical conductivity across the tilt-driven Mott
transition above. Consistent with the small gap ($\Delta_{MH}\simeq 0.1$~eV) Mott
insulator in Fig.~\ref{fig2} for $\theta=160$, the optical conductivity,
$\sigma(\omega)$, shows an onset from $\Omega=2\Delta_{MH}\simeq 0.2$~eV, along
with a mid-infrared feature and incoherent Hubbard band features at higher energy.
The multiple Hubbard peaks arise from inter-orbital ($U',\, J_{H}$) and intra-orbital
($U$) processes across the Hubbard bands in the Mott insulator. With reduction of
the GdFeO$_{3}$ tilt, onset of (bad) metallicity is reflected in $\sigma(\omega)$
as a large scale transfer of high-energy spectral weight to low energies, on the scale
of more than $4.0$~eV, a characteristic fingerprint of Mottness. Additionally, no
quasi-coherent Drude (LFL) peak is resolved in the spectra: instead, a low-energy
pseudogap feature (with a mid-infrared hump) is clearly seen for $\theta=170$.
This evolves into an anomalous continuum with a slow fall-off in energy (and smeared
mid-infrared feature) for $\theta=180$, signalling increasing tendency toward low-energy
coherence as the GdFeO$_{3}$ tilt is reduced. The marked absence of a quasi-coherent Drude response at low energy is a clear manifestation of the wipe-out of the lattice (LFL) coherence scale in the selective-Mott metal in DMFT.  Thus, analysis of optical response of half-metallic and multi-orbital-based TMOs could be used to unearth the multi-orbiton continuum at intermediate-but-low T.  Optical studies of TMO thin films
the $U/t$ ratio in $H$) could confirm this trend in heterostructures: we predict
that varying the A-cation in ABO$_{3}$ films will show the evolution of $\sigma(\omega)$
from Mott insulating to bad-metallic as the A-cation size is increased, e.g,
changing the rare-earth (R) ion in RNiO$_{3}$ or doped RMnO$_{3}$ films from R=Ho to R=La across the
lanthanide series.

Our work suggests that selective-Mottness, bad-metallicity and novel fractionalised excitations,
hitherto features of theories for certain $f$-electron metals and cuprates, might be found ``across
the board'' as orbital-FL$^{*}$ states in correlated TMOs.  In particular, it is known that ARPES lineshapes 
and transport data for bilayer manganites show very ``strange'' features:  ARPES lineshapes are anomalously
broad, as in near-optimally doped cuprates, especially at temperatures higher than the orbital and magnetic ordering scales, and momentum-space (nodal-versus-anti-nodal) differentiation persists even above $T_{c}^{FM}$.  It would be tempting to try and link such ``strange'' features to our proposed OFL$^{*}$ state. 
However, in spite of very close similarities to the FL$^{*}$ idea in Kondo-RKKY models, more formal
field-theoretic work is needed to tease out the fractionalised orbitons in multi-orbital systems showing OSMT.  This is beyond scope of the present work, and is left for future work.   
 
\section { Effects of strain}
 
The above results can now be readily used to gain qualitative insights into the effects
of epitaxial strain on conduction states (metal-insulator, orbital-magnetic order) in
TMO films grown on suitable substrates, e.g, LaNiO$_{3}$/LaAlO$_{3}$ (LNO/LAO)
superlattice grown on a TiO$_{2}$-terminated (001-face) SrTiO$_{3}$ substrate.  We also qualitatively
argue how competing (novel) ordered states may arise upon appropriate ``strain engineering'' in TMO
thin films.

  We begin by adapting our DMFT results and insights to a general model that is expected to hold for the system of an interface between a transition-metal oxide (manganites or LaNiO$_{3}$, for example) on another transition-metal oxide substrate with ABO$_{3}$ structure but with slightly different lattice parameters (e.g, SrTiO$_{3}$).  In such a simplified but very effective model~\cite{lepetit}, the essentials can be recast in terms of energy balance at the interface between competing energies:
 
$(i)$ the film cell parameters are constrained to fit those of the substrate.  Since this implies bond elongation of film cell constants, it relaxes slowly.  This constraint produces a unit-cell volume constraint: $V_{film}=V_{bulk}$, that appears as an elastic term in the free energy, given by $F_{el}=\frac{V}{2\kappa}(\frac{\delta V}{V})^{2}$.  For LSMO on STO, this contribution is $O(0.015)$~eV~\cite{lepetit}.

$(ii)$  Due to epitaxial constraints (compressive or tensile strain affecting the inplanar hoppings, and the GdFeO$_{3}$ tilts at the interface), the electronic structure of the film undergoes drastic reconstruction, especially so in conditions realizing an OSMT, as described above.  Let us consider the case where the hopping integral between the film and the substrate orbitals is affected by the renormalization of the electronic structre of the film.  For example, for the LSMO/STO interface, this integral corresponds to the matrix element

\begin{equation}
t_{\perp}\simeq -\frac{t_{Mn,d_{z^{2}},Ti,d_{z^{2}}}^{2}}{\epsilon_{Ti,d_{z}^{2}}-\epsilon_{Mn,d_{z}^{2}}}
\end{equation}

where $$t_{Mn,d_{z^{2}},Ti,d_{z^{2}}}\simeq \frac{\langle d_{Mn,z^{2}}|p_{{O,z}}\rangle\langle p_{O,z}|d_{Ti,z^{2}}\rangle}{\epsilon_{d}-\epsilon_{p}}$$.  
Owing to the short metal-oxygen distance ($~\simeq 1.95A$), within LDA, this turns out to be $O(0.5)$~eV in a best qualitative estimate, while $(\epsilon_{d}-\epsilon_{p})\simeq 1.5-2.0$~eV. 
Plugging these into $t_{\perp}$ yields $t_{\perp}\simeq 0.125$~eV.  Near an OSMT, however, there is a further drastic many-body renormalization due to the very small $z_{FL}$.  If we take typical values for $z_{FL}=0.05-0.1$, the renormalized $t_{\perp}$ in a strongly correlated situation is now scaled down to $0.0125$~eV
(this is in contrast to the earlier estimate~\cite{lepetit}, which gave a much larger $t_{\perp} \simeq O(0.1-0.5)$~eV).

   These estimates now allow us to use this simple qualitative model to estimate the fate of the structural changes at the interface.  This is because one realizes the interfacial structure with $c<a$ if the elastic energy wins over the electronic delocalization energy, while another structure with $c>a$ obtains when the electronic delocalization energy exceeds the elastic energy.  It is clear from (i),(ii) above that this is a rather delicate balancing act in interfaces made from strongly correlated systems.  In particular, small changes in epitaxial constraints via strain tuning mentioned in the introduction can now easily tilt the balance between these two energies, leading to rather pronounced structural-cum-electronic changes at the interface.  We detail upon this aspect below.
   
\noindent (i) Tensile strain ($e_{xy}>0$) reduces the hopping via $t(e_{xy})=t(1-2e_{xy})$.
With increasing tensile strain, the correlated metal can thus be driven continuously to a Mott
insulator. Further, since T-strain stabilizes the $x^{2}-y^{2}$ relative to the $3z^{2}-r^{2}$
orbital, it increases the tendency toward an effective one-band model in the
less-than-or-equal to the quarter-filled case, with ferro-orbital order and
antiferromagnetic spin order, as in the cuprates.  This is borne out by an explicit
calculation on a related but different model for manganites~\cite{baena}, which shows that 
$e_{xy}>0$ increases the extent of the $A$-phase, reducing those of the FM and CE phases. 
 
  For not-too-large separation between the $x^{2}-y^{2}$ and $3z^{2}-r^{2}$ orbitals, the $d$-wave
interband hybridisation gives rise to a novel possibility: it is conducive for observation of
the exotic circulating current phase proposed by Varma for cuprates~\cite{varma}, but with the
circulating current patterns on three-site triangles connecting two in-plane $O$ sites to the apical
one~\cite{weber,laad-CC} if $J_{H}$ is small. An important difference from the cuprates is that the $3z^{2}-r^{2}$ orbital is
now empty, rather than filled; nevertheless, for not-too-large $\delta=(\epsilon_{3z^{2}-r^{2}}-\epsilon_{x^{2}-y^{2}}$, the $d$-wave hybridisation in the two-band Hubbard model can populate the $3z^{2}-r^{2}$ orbital and still give CC phases~\cite{laad-CC}.
 While this fascinating proposal is presently intensively debated
in the cuprate context, we propose that appropriate strain engineering might reveal its
existence in $e_{g}$-orbital based TMO heterostructures.
This mapping to an effective one-orbital model {\it with} an additional $d$-wave inter-orbital hybridisation
in the tight-binding fit in the $a,b$ basis has further novel consequences.  In particular, multi-orbital,
unconventional $d$-wave nematic order can also readily arise as a particle-hole condensate via a finite
$\Delta_{n}=\sum_{k}($cos$k_{x}-$cos$k_{y})\langle a_{k\sigma}^{\dag}b_{k\sigma}\rangle$.  This will occur in a
way similar to that found for the bilayer ruthenate Sr$_{3}$Ru$_{2}$O$_{7}$~\cite{wu} and for underdoped
Fe-arsenides~\cite{feas-nem}.  Additionally, a $d$-wave superconductor can also result as a particle-particle
condensate with a finite $\Delta_{dsc}=\sum_{k}\gamma_{ab}(k)\langle a_{k\sigma}^{\dag}b_{-k-\sigma}^{\dag}\rangle$,
where $\gamma_{ab}(k)=($cos$k_{x}-$cos$k_{y})$.  Both these ordered states must result from the same microscopic
interaction, and thus must be competing orders.  In the non-LFL metal with infra-red singularities found in DMFT, these ordered states will result from a generalised
Hartree-Fock (GHF) decoupling of the most relevant {\it residual} intersite two-particle interactions (since the
one-electron mixing term ($t_{ab}$) in $H_{2D}$ is irrelevant in the critical metal in the absence of well-defined
LFL quasiparticles).  This residual two particle interaction is analogous to spin-orbital superexchange in
Mott insulating phases, and is ipso-facto justified in the OSMT phase we find~\cite{feas-SC}:
$H_{res}=-(t_{ab}^{2}/(U'+J_{H})\sum_{k,k',\sigma,\sigma'}\gamma_{ab}(k)\gamma_{ab}(k')a_{k\sigma}^{\dag}b_{k\sigma}b_{k'\sigma'}^{\dag}a_{k'\sigma'}$.    
Since this interaction scales like $1/D$ (with $D$ the spatial dimensionality), a GHF decoupling indeed turns out to
be {\it exact} within our DMFT.   
         
\noindent (ii) Compressive strain ($e_{xy}<0$) has exactly the reverse effect - it
increases $t$ and stabilizes the $3z^{2}-r^{2}$ relative to the $x^{2}-y^{2}$ orbital.
Thus, a Mott-insulator to correlated (non-LFL) metal transition can be driven by
increasing compressive strain.  The situation for the ordered states is now more subtle, however.
The stabilisation and preferential occupation of the $3z^{2}-r^{2}$ orbitals implies favoring
 AF order in the bulk.  This goes hand-in-hand with a reversal of the {\it net}     
``splitting'', $\delta=(\epsilon_{3z^{2}-r^{2}}-\epsilon_{x^{2}-y^{2}})$ from positive for
$e_{xy}>0$ to negative for $e_{xy}<0$.  Thus, viewed from the starting point of an unstrained
system ($e_{xy}=0$) with $\delta >0$, C-strain produces, remarkably enough, an effective situation
that corresponds to a system with a {\it negative} inter-orbital charge-transfer (CT) energy.  This identification
allows us to interpret the Mott transition in this case as one driven by ``self-doping'' a negative-CT insulator.
Further, this situation also implies that the $d$-wave (nematic, SC) states discussed for T-strain now become
unfavorable. 
 
  These are concrete examples of how strain engineering can sensitively affect the delicate balance between competing
orders, and produce surprising features not to be found in the bulk.  An exotic scenario would be where the balance between electronic delocalization energy and elastic energy can be tuned by strain engineering.  This would make it conceivable, for example, to realize an unexpected outcome wherein a TMO thin film with $c>a$ can be ``tuned'' into a state with $c<a$ or vice-versa.  In a multi-orbital system, this will drastically reconstruct the interfacial electronic structure, leading to emergence of ordered states not expected in the bulk.  This is because $c>a$ and $c<a$ correspond to ``tensile'' and ``compressive'' strain which, as discussed above, lead to drastically different and novel outcomes for ordering instabilities at the interface via reconstruction of correlated electronic states. 
 
  Other observations may also be qualitatively rationalizable within our model study.  For example, the experimental finding of a reduced conductivity and magnetic ordering scales in manganite
thin films relative to their bulk values~\cite{adamo} is also naturally understood from our results.  This
arises from three factors, all of which conspire to produce an effective reduction in $t(\theta)$:
 
(i) reduction in lattice co-ordination number in the film relative to the bulk reduces screening of
the Hubbard $U,U'$, which increase as a result.
 
(ii) lattice mismatch at the interface reduces the $t(\theta)$ via tolerance factor and strain effects~\cite{baena} effects, which acts like enhanced tilt relative to the bulk GdFeO$_{3}$ tilt.  
 
(iii) larger $(U,U')/t(\theta)$ ratio implies selfconsistent enhancement of the net JT coupling
via increased tendency toward localisation.

\noindent The consequent (self-consistent) reduction of $t(\theta)$ qualitatively explains lowered conductivity
(increased localisation) as well as reduction in magnetic ordering scales, since $T_{N}^{af}\simeq 4t^{2}/(U'-J_{H})$
and for FM, $T_{c}^{fm}\simeq t^{2}/J_{H}$ in the DE limit.  Thus, as qualitatively alluded to in earlier
sections, the structural DE and its interplay with conventional AH-DE is predicted to sensitively affect transport
and magnetic ordering scales in TMO films in ways not always anticipated in the bulk case.
 
  Finally, extension of our DMFT to multilayers requires considration of the $c$-axis hopping ($t_{ab}^{zz}$),
neglected in our $D=2$ modelling.  Baena {\it et al.}~\cite{baena} have considered this aspect in detail in
a related model for manganites.  How these effects will affect physical properties in response to external stimuli
and affect the interplay of competing ordered states within remains an open
issue of great interest in the TMO heterostructure context.
 
\section{ Conclusion}
 
  In conclusion, we have considered the detailed effect of the GdFeO$_{3}$ tilt, ubiquitous in TMO systems.  Using
the two-orbital Hubbard model as a template, we have identified a new feature, namely, that
this tilt acts like a {\it structural} ``reverse double exchange'' in the orbital sector.  It thus acts contrary to
the Anderson-Hasegawa (magnetic) double exchange, and the competition between these two DE processes is shown to
be a generically relevant feature in TMOs.  This additional control parameter is attractive in the sense of not causing
(doping induced) disorder, and can be continuously tuned by suitable choice of substrate and strain engineering.
Adopting a strong correlation view of TMO, this parameter sensitively affects carrier itinerance, inducing phase
transitions between Mott insulators and
incoherent metals with ill-defined LFL quasiparticles.  This breakdown of the LFL metal arises from the
orbital-selective character of the Mott transition, and we argue that this is an orbital-FL$^{*}$ state with
fractionalised orbitons. As a particularly attractive fall-out of our work, suitable strain engineering is
also predicted to generate situations conducive to novel ordered states, whose further exploration is undoubtedly
of great topical interest.  A detailed study of this last aspect is left for future consideration.

\end{document}